\documentclass{mem}
\usepackage{natbib}\usepackage{txfonts}\usepackage{balance}
\usepackage{graphicx}
\idline{00}{000}
\begin{document}
\def\teff{$T\rm_{eff }$}
\def\kms{$\mathrm {km s}^{-1}$}
\def\msun{M$\rm_{\odot}$}

\title{
Interpreting the complex CMDs of the Magellanic Clouds clusters}

   \subtitle{}

\author{
I. Cabrera-Ziri\inst{1,4}
\and S. Martocchia\inst{2} 
\and K. Hollyhead\inst{3}
 \and N. Bastian\inst{2}
 }

\institute{
Harvard-Smithsonian Center for Astrophysics, 60 Garden Street, Cambridge, MA 02138, USA
\and
Astrophysics Research Institute, Liverpool John Moores University, 146 Brownlow Hill, Liverpool L3 5RF, UK
\and
Department of Astronomy, Oskar Klein Centre, Stockholm University, AlbaNova University Centre, SE-106 91 Stockholm, Sweden
\and{Hubble Fellow}
\email{ivan.cabrera@cfa.harvard.edu}
}

\authorrunning{Cabrera-Ziri}

\titlerunning{the CMDs of Magellanic Clouds star clusters}

\abstract{
The Magellanic Clouds host a large population of massive ($>10^4$ \msun) star clusters with ages ranging from a few Myr to ~12 Gyr.  In nearly all cases, close inspection of their CMDs reveals features that deviate from expectations of a classic isochrone. Young ($<2$ Gyr) clusters show extended main sequence turnoffs and in some cases split/dual main sequences. Clusters older than $\sim2$ Gyr show splitting in the red giant branches when viewed in UV filters that are sensitive to abundance variations (in particular nitrogen). A distribution of stellar rotation rates appears to be the cause of the complex features observed in the young and intermediate age clusters, while above $\sim2$ Gyr the features seem to be the same light-element abundance variations as observed in the ancient Galactic globular clusters, a.k.a. ``multiple populations".  Here, we provide an overview of current observations and their interpretations and summarise possible links between all the classes of complexities, regardless of age.

}
\maketitle{}

\section{Introduction}

In the late 1950s Eric Lindsay and Gerald Kron were able to resolve some of the ``nebulous patches" found in the Magellanic Clouds. They realised that a fraction of them were in fact star clusters and pioneered the first systematic studies of the cluster population of the Magellanic Clouds. More than sixty years later since the firsts publications of \cite{Lin56} and \cite{K56}, the study of the cluster population of the Magellanic Clouds is still one of the most exciting and active fields in astronomy.

Today it is well known that the Magellanic Clouds host a large population of massive clusters ($>5\times10^3$ \msun) \citep[e.g.][]{H03,Baum15}. The colour-magnitude diagrams (CMDs) of these massive clusters are perfect laboratories for stellar evolution since most evolutionary phases are well sampled. These clusters suffer from low extinction/reddening and their populations stand out clearly from contaminating foreground/background field stars in the CMDs, unlike clusters in the Galaxy. Furthermore, despite their distance (LMC $\sim50$ kpc and SMC $\sim60$ kpc), the Clouds provide some of the best CMDs of clusters younger than $\sim9$ Gyr.

In 2007, \citeauthor{MBN07} noticed that the CMD of the massive ($\sim10^5$ \msun) $\sim1.7$ Gyr LMC cluster NGC 1846, showed two clear turn-offs. Subsequent photometric surveys, have revealed that many other LMC/SMC clusters show CMDs more complex than what is expected from an isochrone describing a single stellar population. The CMDs of clusters of different ages show different complex features, for example:

\begin{itemize}
\item Young ($<500$ Myr) clusters have been found to have a ``split" or dual main sequences \cite[split-MS -- e.g.][]{D17}.
\item Young and intermediate age ($<2$ Gyr) clusters show multiple or extended main sequence turn-offs \citep[eMSTO - e.g.][]{Mi09}. 
\item Intermediate age and ancient cluster ($> 2$ Gyr) display multiple red giant branch (RGB) sequences on their CMDs \citep[e.g.][]{N17b}. 
\end{itemize}

Current evidence suggests that some of these features are most likely consequences of different phenomena. Here we present a concise account of the evidence and different interpretations of these complex CMDs.

\section{The CMDs of $<2$ Gyr old clusters}

The eMSTOs of intermediate ($\sim1-2$ Gyr) clusters was the first evidence for complex and unexpected features in the CMDs of clusters in the LMC/SMC \citep[cf.][for CMDs of an early survey]{Mi09}. The first studies demonstrated field star contamination nor binaries can fully account for the observed colour spread in the turn-off of these clusters \citep[e.g.][]{G09}, so an unusual effect must be causing the CMDs of these cluster to behave this way.

While originally found in the intermediate age (1-2 Gyr) clusters, it was later found in younger ($\sim10^7-10^8$ yr) clusters as well \citep[e.g.][]{Mi15,C15,B16}. However, in addition to the eMSTO some of the youngest ($\le \sim400$ Myr) clusters also showed a clear split of their main sequence stars \citep[e.g. CMDs from a recent compilation by][]{D17}.

\begin{figure}[]
\resizebox{\hsize}{!}{\includegraphics[clip=true]{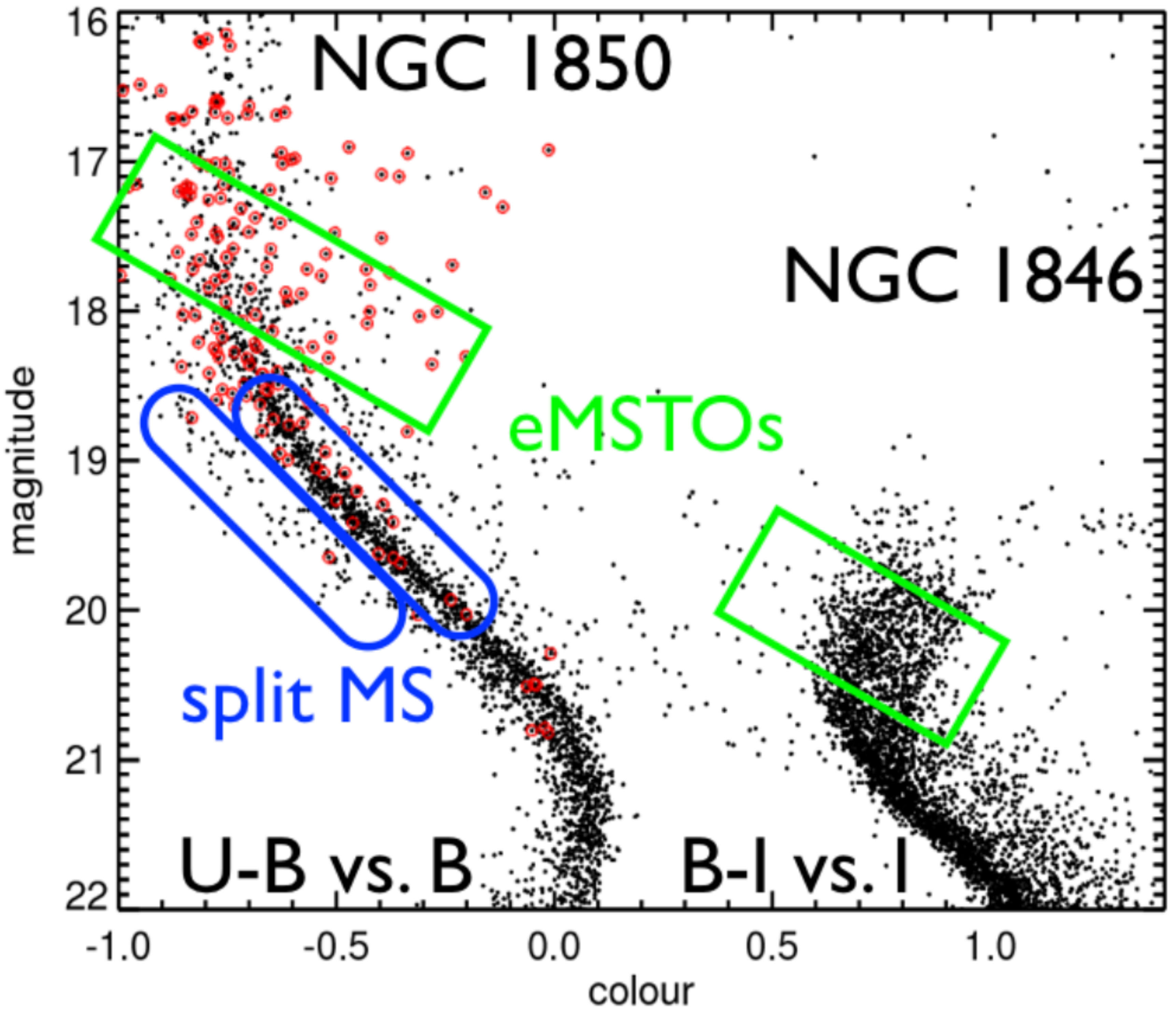}}
\caption{
\footnotesize
CMDs of two young star cluster in the LMC. Left: NGC 1850 ($\sim100$ Myr) displaying an example of a split-MS and eMSTO. Right: NGC 1846 a $\sim1.4$ Gyr cluster with an eMSTO.
}
\label{cmd}
\end{figure}

Figure \ref{cmd} provides an example of the CMDs of two young clusters hosting these complex features.

\subsection{Important observational constraints} 
\label{prop}
The most relevant characteristics that could help us understand the complex CMDs of these young clusters are: 1) that the extent/area of the eMSTO seems to be strongly correlated with cluster age (cf. Fig. \ref{niederhofer}); 2) no eMSTO is found in clusters above an age of $\sim2$ Gyr (cf. Fig. \ref{old}); 3) unlike Galactic GCs, their RGB seem consistent with a single stellar population in CMDs sensitive to abundance variations and 4) some of the $\sim10^8$ yr clusters host large fractions of H$\alpha$ emitters.\footnote{more about points 3) and 4) in \S\ref{MPs}}

For the rest of this section we will outline the basic principle behind the main hypothesis proposed to explain the split-MS/eMSTO phenomenon and we will put them in context of the known properties we have just described.

\subsection{Age spreads} 

The most logical and simple interpretation of the eMSTO phenomenon was that the turn-off of these clusters was the consequence of multiple star formation events.

This interpretation gained strength as it coincided with the popular notion that the star-to-star differences in light elements of globular clusters were product of multiple star formation events \cite[cf.][for a recent review]{C16}. 

Given the scarcity of data about the chemical compositions of the stars in these clusters, during the early days of these scenarios, most of the work was focused to present detailed properties of the young clusters at the time they underwent the multiple star formation events, and describe their subsequent evolution.

These scenarios posit that the young cluster should have been significantly more massive (had larger escape velocities) than it is today. This would have allowed the cluster to retain and accrete gas to fuel subsequent star-formation events. After the formation of the 2nd generation, the cluster would proceed to lose most of its initial stars.
 The cluster then continues evolving passively throughout the rest of its life. \cite{G14} presents one of the most developed scenarios along this line of thought, so we refer it to the reader  interested in a detailed description.

\begin{figure}[]
\resizebox{\hsize}{!}{\includegraphics[clip=true]{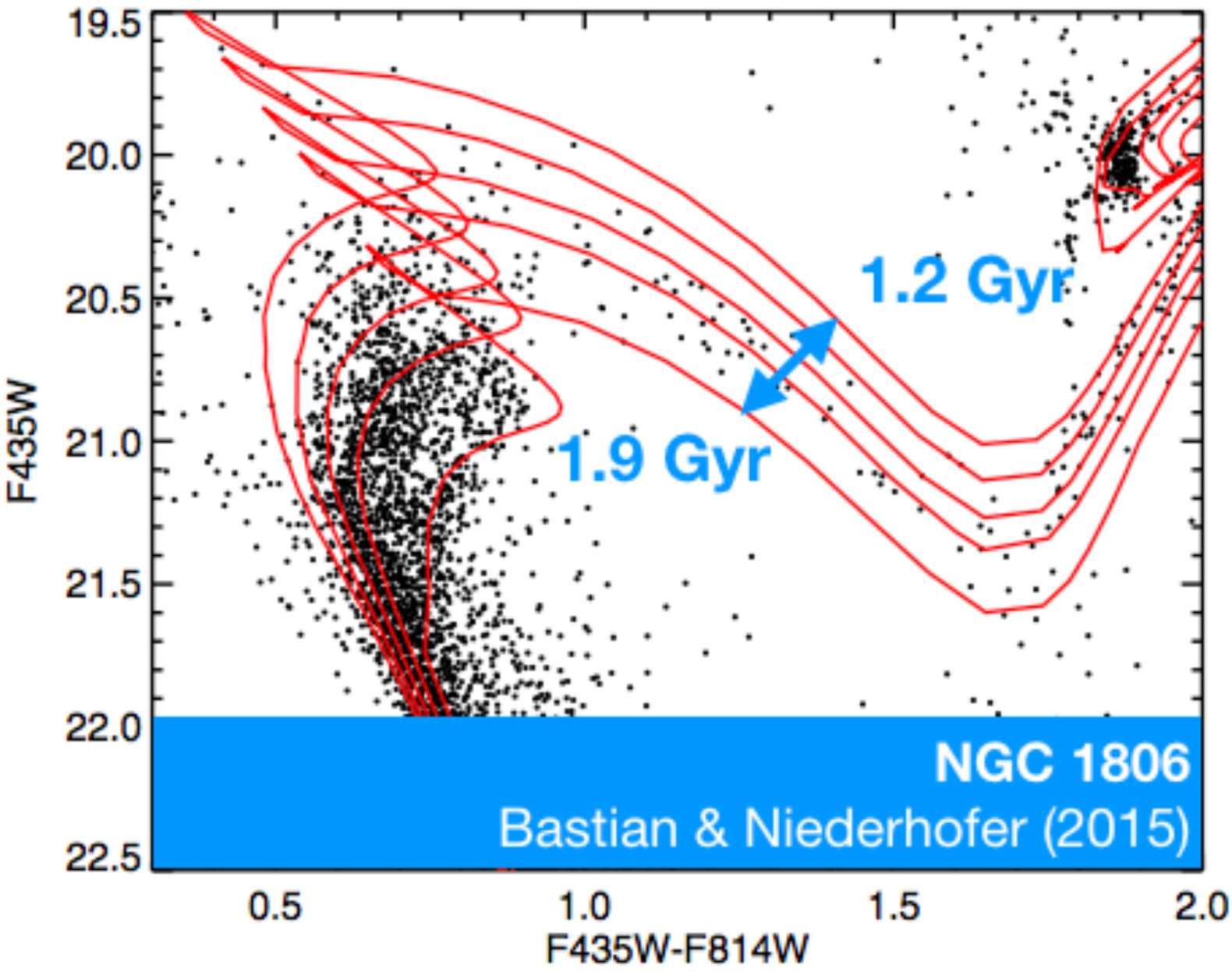}}
\caption{
\footnotesize
Figure from \cite{BN15} showing the inconsistency between the inferred star formation history from the eMSTO and other evolutionary stages like the SGB and RC. Red lines show isochrones for ages between 1.2 and 2 Gyr.
}
\label{bn}
\end{figure}

According to this notion the eMSTO would be successfully reproduced by turn-offs of a wide range of ages. However, this concept has several caveats. For example, different post-main sequence evolutionary stages like the sub-giant branch (SGB) and the red clump (RC) seem incompatibles with the extended star formations histories inferred from the eMSTO \cite[e.g.][cf. Fig. \ref{bn}]{L14,N16}. It also seems that the mass loss suffered by these young clusters in the LMC/SMC have been significantly overestimated \cite[cf.][]{CZ16}.

This scenario does not predict or account for any of the properties listed in \S \ref{prop}.

\subsubsection{Abundance variations} 

\cite{Mi15} and \cite{Mi16} experimented varying the He, C+N+O and [Fe/H] abundances of clusters with split-MS assuming different ages for the populations with different abundances. These attempts failed to reproduce the split-MS satisfactory and suffered from the same caveats (regarding the multiple star formation events) mentioned earlier.

It is worth noting that to date, there is no spectroscopic or photometric evidence suggesting abundance variations in young ($<\sim$2 Gyr) Magellanic Cloud clusters, cf. \cite{Muc08,M17a,M17b}.

\subsection{Variable stars}

\cite{Salinas16} explored the contribution of variable stars to the eMSTO phenomenon. They pointed out that the instability strip in the CMDs of intermediate age clusters overlaps with their turn-off, and as a consequence it is expected to find a population of variable stars in this region (namely $\delta$ Scuti). These authors showed that a colour spread in the MSTO can be introduced by the effect of the changes in brightness of $\delta$ Scuti stars.

Variable stars can account for the lack of eMSTO in older ($\sim$2.5 Gyr) clusters since the turn-off of the older clusters lies outside the instability strip. Since the red clump is also outside the instability strip, this could explain why it is not affected. Furthermore, $\delta$ Scuti can partially account for the colour spread as a function of age (cf. Fig. 4 of \citealt{Salinas16}), however, in this case the predicted changes are much smaller than the observed ones. 

Despite that, stellar variability cannot account for the split-MS, the bimodal turn-off nor the large fraction of H$\alpha$ emitters found in some clusters. For clusters $<\sim$1 Gyr, the turn-off would be bluer than the instability strip, hence the eMSTO of younger clusters is unlikely the effect of variable stars. The incidence of variables stars, and their respective amplitudes needs to be large in order to reproduce the observed properties of eMSTO clusters.

Recent studies of the eMSTO cluster NGC 1846 reported $<60$ $\delta$ Scuti, suggesting that a large incidence of these stars in this kind of cluster in unlikely \citep{Pajkos16}.

\subsection{Fast rotators}

Fast stellar rotation has two main effects that are capable affecting the detectable shape of the turn-off, namely: the internal mixing and viewing angle of the stars. Due to fast rotation the internal mixing increases, which brings more fuel (fresh Hydrogen) into the core of the stars. As a consequence they remain longer on the MS than stars rotating at a ``normal" rate.

\begin{figure*}[t!]
\resizebox{\hsize}{!}{\includegraphics[clip=true]{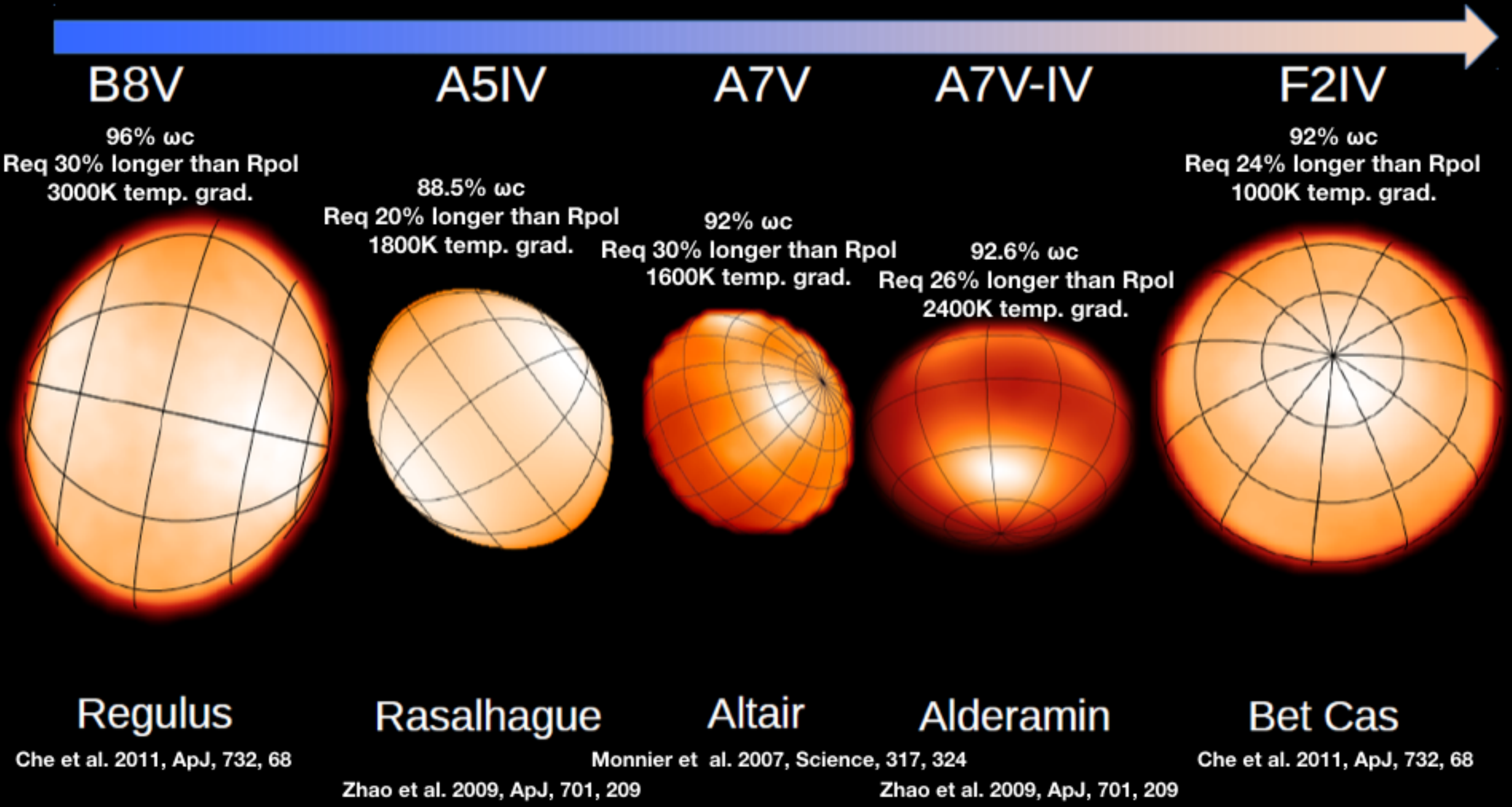}}
\caption{\footnotesize
Images of the stellar discs of fast rotators from the CHARA Array (www.chara.gsu.edu) by John Monnier. For each star we report: their velocity as a percentage of the critical rotation rate $\omega_c$; the difference between their radius at the equator and pole (R$_{eq}$ and R$_{pol}$, respectively); and the temperature gradient between the pole and equator.}
\label{chara}
\end{figure*}

Fast rotators can be quite deformed due to their extreme centrifugal forces (cf. Fig. \ref{chara}). This would produce lower effective temperatures at their equators than their poles. As a consequence, the same star will show different temperatures depending on the inclination angle we see them.

It is possible to reproduce the eMSTO by invoking populations of rapid rotators with different rotation rates and viewing angles, in a cluster where all the stars share the same age and metallicity, e.g.  \cite{BdM09,BH15}. Moreover, the split-MS can be accurately modelled by coeval populations of stars (of the same metallicity) with two distinct rotation rates, where one population is essentially none-rotating while the other one rotates at almost critical rate\footnote{The break-up velocity, critical velocity, or critical rotation of a fast rotating star is the surface velocity at which the centrifugal force just matches the force of gravity. Beyond this point, the star would begin to eject matter from its surface \citep{Mae08}.} \citep[e.g.][]{D15}.

\begin{figure}[]
\resizebox{\hsize}{!}{\includegraphics[clip=true]{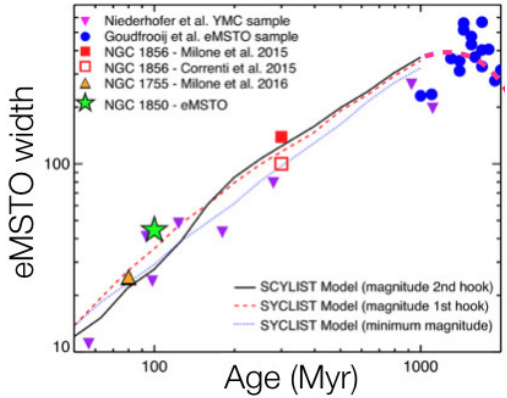}}
\caption{
\footnotesize
Figure from \cite{N15}, updated with data from recent publications. Different lines show the prediction of the eMSTO width as a function of age for different models. The symbols denote the eMSTO width of different cluster.
}
\label{niederhofer}
\end{figure}

The post main sequence features like the SGB and RC are expected to be compact in fast rotating populations \citep[cf.][]{N15}. Additionally, stellar models naturally predict that the width of the turn-off of a coeval population with a distribution of rotation velocities changes as a function of age \citep[cf.][see Fig. \ref{niederhofer} for a comparison with data]{BH15}.

With that said, the fraction of fast rotators found in stars in the field or in open cluster is significantly lower \cite[$<10\%$, e.g.][]{MG05} compared to the fraction of fast rotators (most of the cluster stars) required by models to reproduce the complex CMDs of young and intermediate age clusters. \emph{Are there large fractions of critically rotating stars in such clusters?}

If large populations of fast rotating stars were to be responsible of the split-MS of these young clusters, the clusters should also host large fraction of Be stars. Be stars are main sequence B stars rotating near critical rate. They have decretion discs which (sometimes) are ionised by the host star. Their SED is the result of a composite of: photospheric emission, disc emission of reprocessed radiation and disc absorption \cite[cf. ][for a comprehensive review on Be stars]{R13}. 

\cite{B17} used stars with excess H$\alpha$ emission as a proxy of the Be population\footnote{Since the Be phenomenon is intermittent there is a fraction of the population rapidly rotating that will not be detectable in  H$\alpha$ emission. Consequently the observed fraction of H$\alpha$ emitters represents a lower limit to the actual number of fast rotators within the clusters.}. They reported large fractions ($\sim30 - 60\%$) of Be stars in the turn-off of NGC 1850 ($\sim80$ Myr) and NGC 1856  ($\sim280$ Myr), arguing in favour of the interpretation that the split-MS and eMSTO of clusters is due to the effect of fast rotators. 

\cite{Dup17} studied turn-off stars of NGC 1866 ($\sim200$ Myr) providing the first spectroscopic confirmation of Be stars (near-critically rotating H$\alpha$ emitters) in a young LMC cluster. Furthermore, they were able to derive actual rotation rates from the profiles of absorption lines. They concluded that this cluster hosted a population of fast rotators ($ \gtrsim150$ km/s) and another one of slow rotators.
The \citeauthor{Dup17} observations represent strong evidence for a rapid stellar rotation as the origin of the split-MS and eMSTO in young Magellanic Cloud's clusters.

\section{The CMDs of $>2$ Gyr old clusters}
\label{MPs}

\begin{figure*}[t!]
\resizebox{\hsize}{!}{\includegraphics[clip=true]{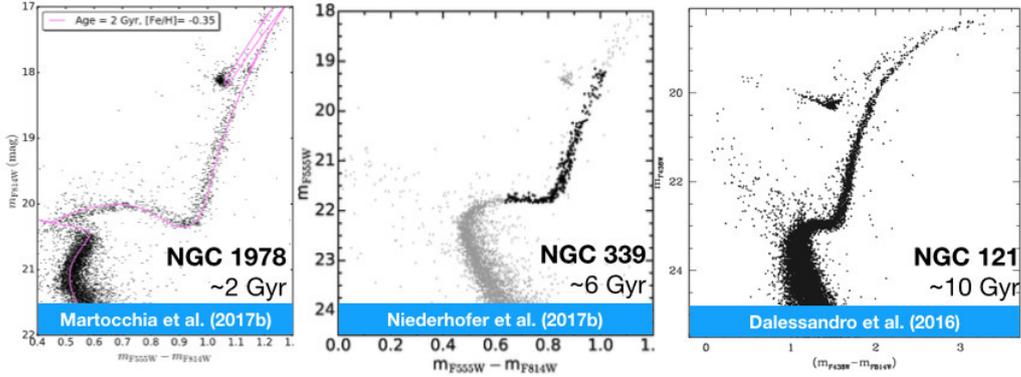}}
\caption{\footnotesize
The CMDs of three old clusters in the Magellanic Clouds. In a conventional optical CMDs, these clusters are well described by a simple isochrone. }
\label{old}
\end{figure*}

Contrary to what is found on the younger clusters, the optical CMDs of clusters older than $\sim2$ Gyr is well described by a simple isochrone \citep[e.g.][see Fig. \ref{old}]{Dal16,N17b}. These clusters show no evidence of an eMSTO or a split-MS. 

Old Galactic globular clusters ($\sim10$ Gyr), are known to host star-to-star abundance variations of He and some light elements (e.g. C, N, O, Na, Al, Mg, etc.). These abundance variations have been confirmed to be present as well in old LMC globulars \citep[cf.][]{Muc09}.

We can use the CMDs of these clusters to infer differences in the chemical abundances between the subpopulations, in the same way it is done in Galactic globulars \citep[e.g.][]{Mi17}.

\subsection{Horizontal branch morphology}

\begin{figure}[]
\resizebox{\hsize}{!}{\includegraphics[clip=true]{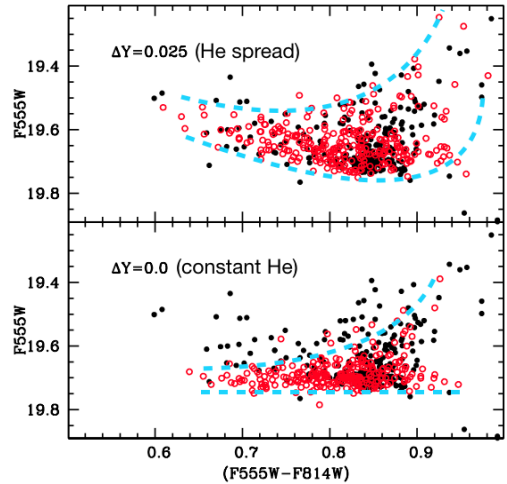}}
\caption{
\footnotesize
Red HB stars of NGC 121 are shown as black dots. The red open circles show the results of simulations modelling the HB morphology with different initial helium mass fractions.  The observed data is better reproduce by models that account for a He spread in this cluster ($Y$ 0.248 to 0.273 top panel), than a population with a constant He abundance (bottom panel). We refer to \cite{N17a} and references therein for more details.
}
\label{hb}
\end{figure}

The wedged shape of the red horizontal branch (HB) in the optical CMDs of globular clusters is a consequence of a range of He abundances of the cluster's stars \citep[e.g.][]{Sal16}. This feature, cannot be reproduced by stellar populations that do not account for a significant star-to-star variation of the initial He abundances. With this in mind \cite{N17a} modelled the HB of the $\sim10$ Gyr SMC cluster NGC 121 and concluded that it's morphology is consistent with a range of initial He mass fraction ($\Delta Y$) of 0.025 (cf. Fig. \ref{hb}).

Unfortunately, no modelling of the HB and the subsequent He abundance spread of clusters with ages between $\sim2-8$ Gyr have been undertaken so far. These studies are necessary to understand the evolution (if present) of He abundance variations.

\subsection{Red giant branch splits}

\begin{figure}[]
\resizebox{\hsize}{!}{\includegraphics[clip=true]{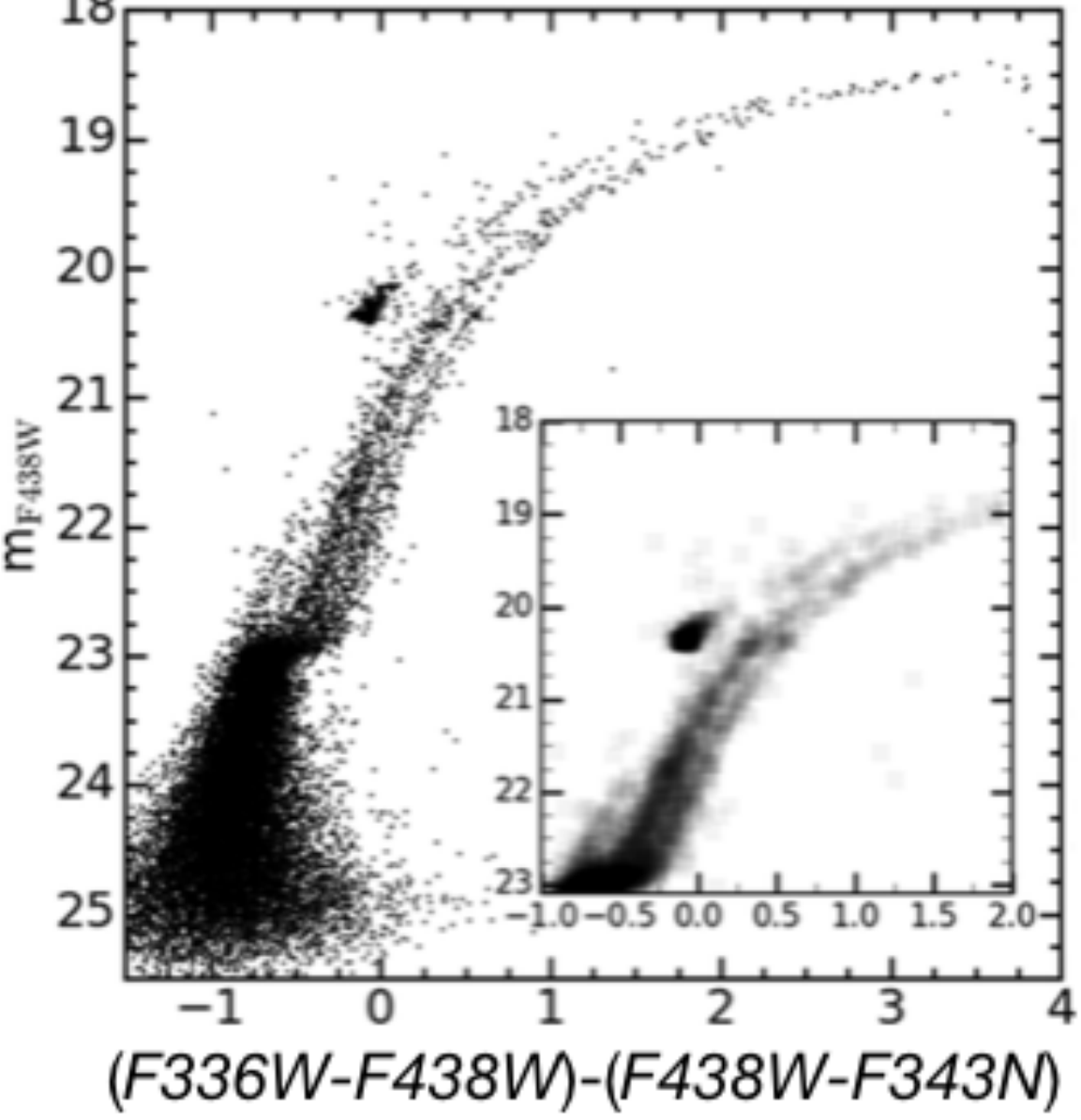}}
\caption{
\footnotesize
CMD of NGC 121 ($\sim10$ Gyr). The pseudo-colour $(F336W - F438W)-(F438W -F343N)$ is a good proxy of N-enrichment. This cluster shows a clear split along the RGB (see inlay). Figure from \cite{N17a}.
}
\label{121}
\end{figure}

Different levels of N-enhancement/C- and O-depletion between subpopulations in GCs can be distinguished with precise photometric measurements using filters containing strong molecular bands, like the OH-, NH-, CN- and CH-bands in the ultraviolet and blue spectrum ($\sim 2500-4500$ \AA~ e.g. Piotto et al., 2015) of RGB stars. The resulting CMDs with photometry focused in this spectral region, maximises the difference between subpopulations, which can be seen as a splitting/spreads in the RGBs (e.g. Fig. \ref{121}).

\begin{figure}[]
\resizebox{\hsize}{!}{\includegraphics[clip=true]{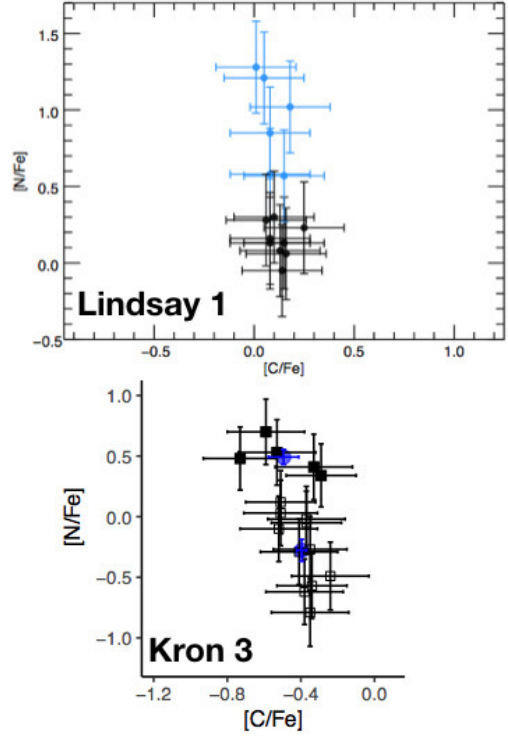}}
\caption{
\footnotesize
N- and C-abundances of Lindsay 1 and Kron 3 stars. The spread in [N/Fe] is clear however, the [C/Fe] remains relatively constant in both clusters. The N-rich and N-poor
stars are members of different sequences in the RGB of Lindsay 1, cf. \cite{N17b}. This information is not available yet for Kron 3.}
\label{cn}
\end{figure}

\begin{figure}[]
\resizebox{\hsize}{!}{\includegraphics[clip=true]{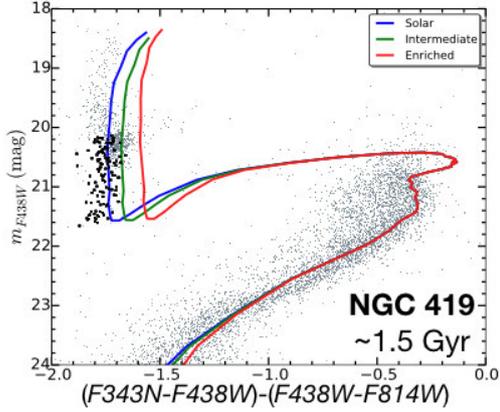}}
\caption{
\footnotesize
The CMD of young clusters $\sim2$ Gyr like NGC 419 suggest homogeneous abundances in the RGB stars.}
\label{419}
\end{figure}

The split of the RGB has been found in the Magellanic Cloud clusters: NGC 121, Lindsay 1, NGC 339, NGC 416 and NGC 1978 \citep[cf.][]{Dal16,N17a,N17b,M17b}. Spectroscopic studies have confirmed the presence of N abundance variations in Lindsay 1 ($\sim7.5$ Gyr) and Kron 3 ($\sim6.5$ Gyr) \cite[cf.][ see Fig. \ref{cn}]{H17a,H17b}. However, no evident spread/split in the RGBs is detected in the $<2$ Gyr clusters analysed to date \cite[cf.][see Fig. \ref{419}]{M17a,M17b}.

In the study of multiple stellar populations of globular clusters, photometric surveys have the advantage over spectroscopic studies in terms of the sample size (even with the latest multiplexing capabilities of the current 8m class spectrographs). The analysis of the CMDs of NGC 1978 and  NGC 121, have revealed the presence of abundance variations in these clusters that were not possible to detect in the small samples of previous spectroscopic studies \citep[cf.][]{Muc08,Dal16}.

\subsection{The onset of abundance variations}

Only with large samples of stars have we been able to detect significant trends of the abundance patters characteristic of old clusters and their basic properties. For example, we now know that the fraction of stars with anomalous abundances correlates with the cluster mass; similarly there is also a correlation between the mass of the cluster and the severity the abundance variations  \citep[cf.][]{Mi17}.

\begin{figure}[]
\resizebox{\hsize}{!}{\includegraphics[clip=true]{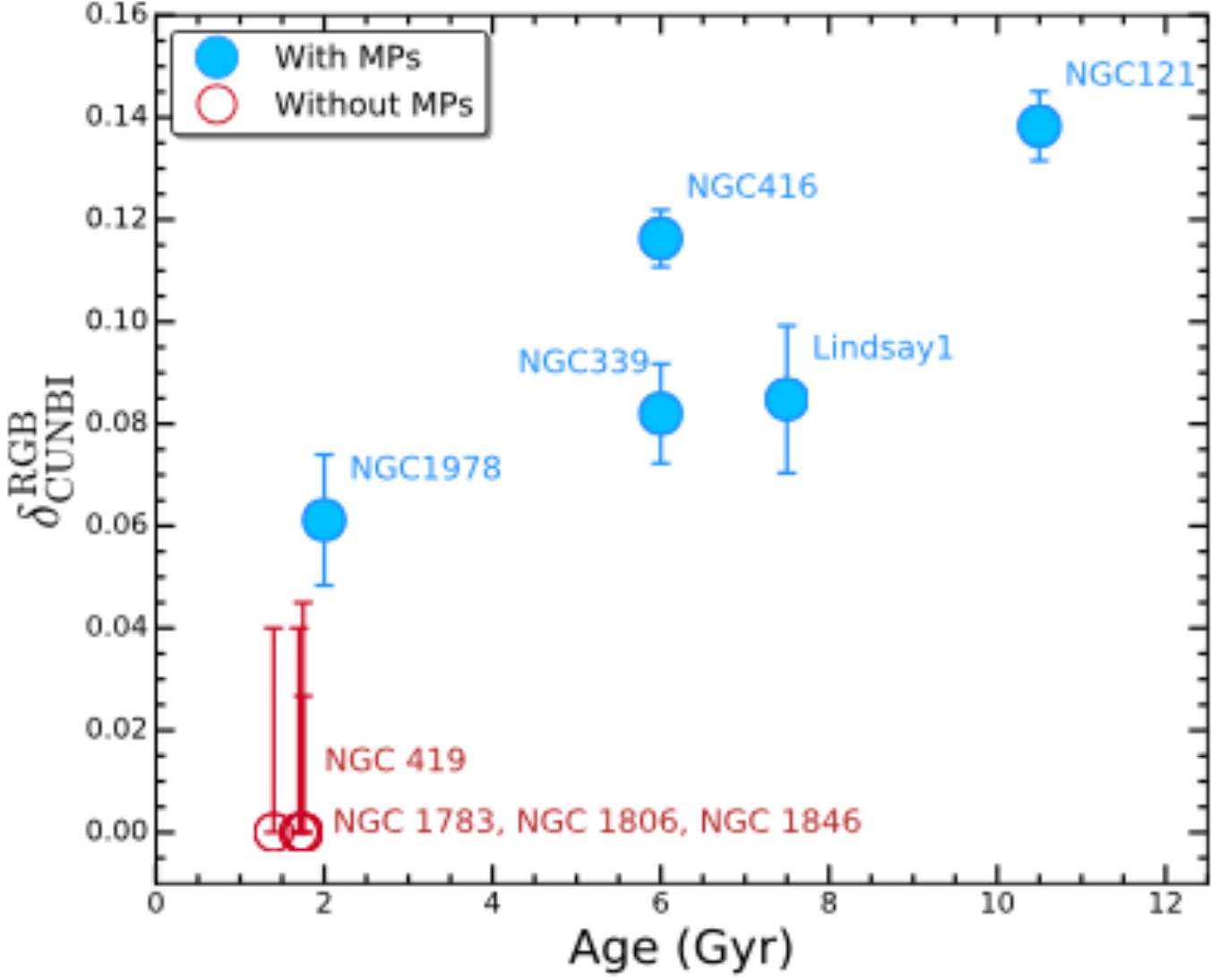}}
\caption{
\footnotesize
Spread in the pseudo-colour RGB in the $(F343N -F438W)-(F438W -F814W)$ as a function of cluster age. This pseudo-colour is mostly sensitive to variations in N-abundances. From this \cite{M17b} inferred that the N-variations seems to be larger in older clusters than in younger ones.}
\label{onset}
\end{figure}

A recent survey of Magellanic Cloud clusters revealed that the width of the RGB in the pseudo-colour $(F343N -F438W)-(F438W -F814W)$ (a proxy of N-enrichement) seems to correlate with the age of the cluster \cite[cf. Fig. \ref{onset} -- ][]{M17b}. In other words, the abundance variations are found to be more severe in older clusters than the younger ones. Current photometric uncertainties place the onset of a evident N-spread in the clusters of the \citeauthor{M17b} sample to be $\sim2$ Gyr. This fresh evidence together with the other correlations reported in \cite{Mi17}, could be critical pieces of the puzzle of the origin of the abundance variations in star clusters.

Furthermore, the upper limit on the possible age spreads present in the youngest cluster where a split RGB have been found are of the order of $6\pm17$ Myr (Martocchia et al. submitted). This leaves little wiggle room for globular cluster formation scenarios that propose multiple star formation events for the presence of star-to-star abundance variations in star clusters. Add to this the latest constraints on the nucleosynthetic sources for the chemical anomalies \citep[i.e.][]{B15,P17} and it seems that the current scenarios for the formation of globular clusters and their characteristic abundance variations are in the need for a deep overhaul.

\section{Summary}

Our proximity to the Magellanic Clouds allow us to resolve their stellar clusters in their constituent stars for spectroscopic and photometric analysis. The CMDs of this cluster population represent snapshots of the process of stellar evolution in dense environments spanning a wide range of ages (few Myr to $\sim12$ Gyr). This opportunity is not available in the Galaxy nor in any of our other Galactic neighbours.

The Magellanic Cloud clusters have inspired several theories --and have served as testbeds-- for the formation and evolution of clusters. Here we have focused on reviewing the complex CMDs of young and old clusters and give an account of the most popular interpretations of the last decade.

The split-MS and eMSTO found in young ($<2$ Gyr) cluster have been the centre of an active debate regarding their origin. Several interpretations have ben proposed to explain the origin of  the CMDs of these clusters, including: age spreads, abundance variations, large populations of variable stars and/or a wide range of stellar rotation. The strength of the evidence supporting each these hypothesis varies significantly. However, the hypothesis that stellar rotation 
is responsible for these phenomena has a predictive character lacking in any of the other alternatives. Furthermore, results from the last year confirmed the presence of the populations rapidly rotating star in these young clusters in excellent agreement with the expectations from this scenario.

On the other hand, the old ($>2$ Gyr) clusters from the Magellanic Clouds seem to have provided an unexpected clue to the long-standing problem that is the origin of the star-to-star abundance variations in globular clusters. The RGB split due to differences in the abundances in old clusters appears to be larger in older clusters, suggesting that the abundance variations get stronger with age. Only by analysing a wide range of ages (not present in the Galactic cluster population) it was possible to reveal this unknown effect that could prove to be key to solve the mystery of the light element abundance variations in star clusters.

The behaviour of the abundance variations with ``global" properties of clusters like their age and mass, must be integrated to the next generation of scenarios aimed at describing this phenomenon.

Having said this, there are still some issues that do not have a satisfactory explanation:

\begin{itemize}
\item Why the stars in clusters seem to be rotating significantly faster and in larger proportion than field stars?
\item What determines the distribution of the inferred rotation rates in young clusters?
\item Is the age where we stop seeing the effects of stellar rotation (e.g. eMSTO), coincidentally the same age were we start seeing the effects of star-to-star abundance within clusters?
\end{itemize}

The answer to the last question will be eagerly pursued, and is likely to become one of the most enticing problems in the studies of stellar evolution during the next years.

\begin{acknowledgements}
Support for this work was provided by NASA through Hubble Fellowship grant \# HST-HF2-51387.001-A awarded by the Space Telescope Science Institute, which is operated by the Association of Universities for Research in Astronomy, Inc., for NASA, under contract NAS5-26555. N.B. gratefully acknowledges financial support from the Royal Society (University Research Fellowship) and the European Research Council (ERC-CoG-646928-Multi-Pop).
 \end{acknowledgements}

\bibliographystyle{aa}

\end{document}